\documentclass[%
 article, twocolumn,
nofootinbib,
 amsmath,amssymb,
 aps
]{revtex4-1}
\usepackage{braket}
\usepackage{hyperref}
\hypersetup{ hidelinks, }
\usepackage{graphicx}
\usepackage{dcolumn}
\usepackage{bm}
\newtheorem{theorem}{Theorem}

\begin{document}


\title{Contextual bound states for qudit magic state distillation}

\author{Shiroman Prakash}
\email{sprakash@dei.ac.in}
\author{Aashi Gupta}
\affiliation{%
 Dayalbagh Educational Institute, \\
 Dayalbagh, Agra, India.
}%

\date{January 3, 2020}
\begin{abstract}
 Identifying necessary and sufficient conditions for universal quantum computing is a long-standing open problem for which  contextuality is, perhaps, the only promising solution [Howard \textit{et al.} Nature
510, 351 (2014)]. To justify this conjecture, Howard \textit{et al.} showed that contextuality is equivalent to Wigner negativity for qudits, and is therefore necessary and possibly sufficient for qudit magic state distillation. Here, we reformulate magic state distillation in the language of discrete phase space, to show that any finite qudit magic state distillation routine contains bound states outside the Wigner polytope -- these states exhibit contextuality but are useless for magic state distillation.
\end{abstract}
\pacs{Valid PACS appear here}

\maketitle

\def\aj{\ref@jnl{AJ}}                   
\def\actaa{\ref@jnl{Acta Astron.}}      
\def\araa{\ref@jnl{ARA\&A}}             
\def\apj{\ref@jnl{ApJ}}                 
\def\apjl{\ref@jnl{ApJ}}                
\def\apjs{\ref@jnl{ApJS}}               
\def\ao{\ref@jnl{Appl.~Opt.}}           
\def\apss{\ref@jnl{Ap\&SS}}             
\def\aap{\ref@jnl{A\&A}}                
\def\aapr{\ref@jnl{A\&A~Rev.}}          
\def\aaps{\ref@jnl{A\&AS}}              
\def\azh{\ref@jnl{AZh}}                 
\def\baas{\ref@jnl{BAAS}}               
\def\bac{\ref@jnl{Bull. astr. Inst. Czechosl.}}
\def\caa{\ref@jnl{Chinese Astron. Astrophys.}}
\def\cjaa{\ref@jnl{Chinese J. Astron. Astrophys.}}
\def\icarus{\ref@jnl{Icarus}}           
\def\jcap{\ref@jnl{J. Cosmology Astropart. Phys.}}
\def\jrasc{\ref@jnl{JRASC}}             
\def\memras{\ref@jnl{MmRAS}}            
\def\mnras{\ref@jnl{MNRAS}}             
\def\na{\ref@jnl{New A}}                
\def\nar{\ref@jnl{New A Rev.}}          
\def\pra{\ref@jnl{Phys.~Rev.~A}}        
\def\prb{\ref@jnl{Phys.~Rev.~B}}        
\def\prc{\ref@jnl{Phys.~Rev.~C}}        
\def\prd{\ref@jnl{Phys.~Rev.~D}}        
\def\pre{\ref@jnl{Phys.~Rev.~E}}        
\def\prl{\ref@jnl{Phys.~Rev.~Lett.}}    
\def\pasa{\ref@jnl{PASA}}               
\def\pasp{\ref@jnl{PASP}}               
\def\pasj{\ref@jnl{PASJ}}               
\def\rmxaa{\ref@jnl{Rev. Mexicana Astron. Astrofis.}}%
\def\qjras{\ref@jnl{QJRAS}}             
\def\skytel{\ref@jnl{S\&T}}             
\def\solphys{\ref@jnl{Sol.~Phys.}}      
\def\sovast{\ref@jnl{Soviet~Ast.}}      
\def\ssr{\ref@jnl{Space~Sci.~Rev.}}     
\def\zap{\ref@jnl{ZAp}}                 
\def\nat{{Nature}}              
\def\iaucirc{\ref@jnl{IAU~Circ.}}       
\def\aplett{\ref@jnl{Astrophys.~Lett.}} 
\def\apspr{\ref@jnl{Astrophys.~Space~Phys.~Res.}}
\def\bain{\ref@jnl{Bull.~Astron.~Inst.~Netherlands}} 
\def\fcp{\ref@jnl{Fund.~Cosmic~Phys.}}  
\def\gca{\ref@jnl{Geochim.~Cosmochim.~Acta}}   
\def\grl{\ref@jnl{Geophys.~Res.~Lett.}} 
\def\jcp{\ref@jnl{J.~Chem.~Phys.}}      
\def\jgr{\ref@jnl{J.~Geophys.~Res.}}    
\def\jqsrt{\ref@jnl{J.~Quant.~Spec.~Radiat.~Transf.}}
\def\memsai{\ref@jnl{Mem.~Soc.~Astron.~Italiana}}
\def\nphysa{\ref@jnl{Nucl.~Phys.~A}}   
\def\physrep{\ref@jnl{Phys.~Rep.}}   
\def\physscr{\ref@jnl{Phys.~Scr}}   
\def\planss{\ref@jnl{Planet.~Space~Sci.}}   
\def\procspie{\ref@jnl{Proc.~SPIE}}   

\let\astap=\aap
\let\apjlett=\apjl
\let\apjsupp=\apjs
\let\applopt=\ao

\section{Introduction and Summary of Results}

Magic state distillation \cite{MSD}, the leading approach to fault-tolerant quantum computation, has been used by Howard \textit{et al.} \cite{nature} to identify \textit{contextuality} \cite{bell, kochen-specker} as a ``necessary and possibly sufficient'' resource for universal quantum computation. This is a  satisfying connection, because contextuality provides a precise operational distinction \cite{Mermin, Spekkens, Abramsky_2011, csw, Acin2015,DPS2} between quantum and classical mechanics and acts as a natural obstruction to classical simulation.


The magic state model \cite{MSD} is based on the observation that the requirement of fault-tolerance naturally imposes strong limitations on the abilities of a quantum computer. The canonical fault-tolerant quantum computer is only able to implement  Clifford unitaries \cite{PhysRevA.57.127}, and initialize and measure qubits or qudits in the computational basis. Via the Gottesman-Knill theorem \cite{gottesman1998heisenberg, PhysRevA.70.052328}-- or discrete phase space for qudits of odd-prime dimension \cite{Veitch_2012} -- a quantum computer subject to these limitations can be efficiently classically simulated. Hence we need to supplement our fault-tolerant quantum computer with an extra ingredient, which, in the magic state distillation model, is the ability to prepare certain non-stabilizer states known as magic states. Using magic states one can implement a non-Clifford gate, such as the $\pi/8$ gate \cite{ACB, HowardVala, PhysRevA.98.032304}, via state-injection, to achieve universal quantum computation. 

By assumption, our hardware can only prepare magic states with low fidelity. Crucially, however, it is possible to produce a magic state with arbitrarily high fidelity, starting from many low-fidelity magic states using  protocols consisting of only Clifford operations and stabilizer measurements. A variety of protocols exist (e.g., \cite{MSD, knill, Reichardt2005, reichardt2009quantum, ACB, CampbellAnwarBrowne, Howard}). For any given protocol, there exists a minimum threshold fidelity required of input states for the protocol to be successful. Input states not meeting this requirement are known as \textit{bound states} \cite{bound}, and are useless for producing pure magic states using the given protocol.

\begin{figure}
\centering
\includegraphics[width=3.4in]{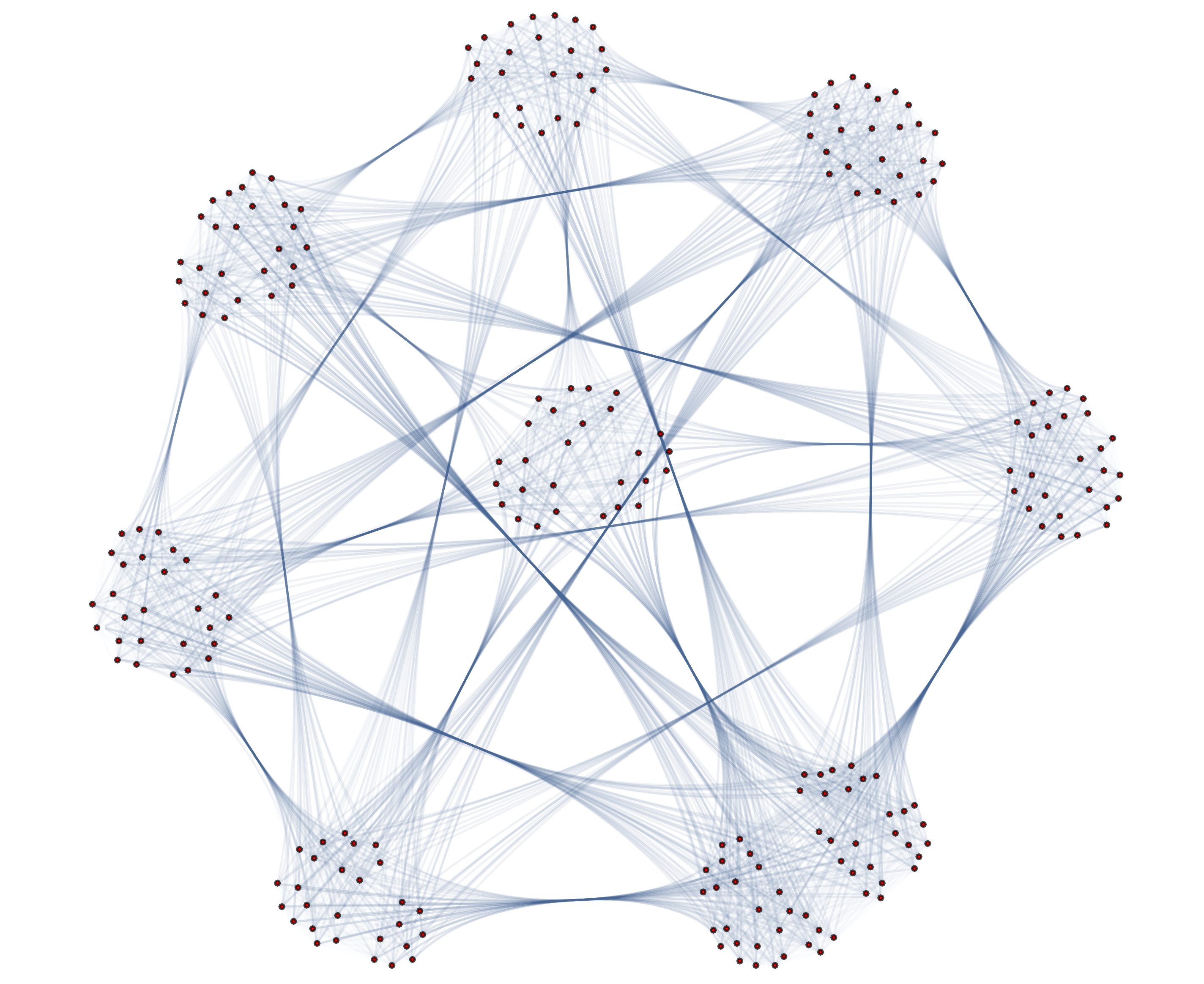}
\caption{\label{exgraph} \textbf{The exclusivity graph \cite{csw} constructed by Howard \textit{et al.} \cite{nature} that serves as a witness for contextuality, as described in section \ref{cont}.}}
\end{figure}

Classical simulability places a theoretical limit for the threshold fidelity of any magic-state distillation routine. Clifford operations on states in the Wigner polytope -- the region of state space consisting of states with positive Wigner function \cite{Wootters1987, PhysRevA.70.062101, Gross} -- are classically simulable, hence any potentially useful input states for magic state distillation must lie outside this polytope \cite{Veitch_2012, PhysRevLett.109.230503}. \cite{nature} showed that the set of single-qudit states which lie outside the Wigner polytope are precisely those states that display contextuality with respect to stabilizer measurements \cite{Howard_2013} --  witnessed by the exclusivity graph \cite{csw} of Figure \ref{exgraph} -- thus establishing contextuality as a necessary condition for universal quantum computation. 

The argument of \cite{nature} crucially invoked qudits of odd-prime dimension. Attempts to generalize this to qubits, \cite{PhysRevLett.119.120505}, lead to difficulties \cite{PhysRevLett.122.140405}, such as the existence of state-independent contextuality. Because the role of contextuality as a resource is most precise for qudits of odd-prime dimension (which, following \cite{nature}, we refer to simply as qudits), we focus exclusively on this case. 

Here, we attempt to address  whether contextuality is \textit{sufficient} for universal quantum computation. There are two ways to translate this question into a question about magic state distillation. 

\begin{enumerate}
    \item Does there exist a magic state distillation routine for which the onset of contextuality is a sufficient condition for distillation?
    \item For any quantum state exhibiting contextuality, does there exist a magic state distillation scheme for which that state is useful?
\end{enumerate}

 Below, we show that the answer to the first question is no. Our result can be stated as
 \begin{theorem}\label{main} For any $N$, there exist qudit states that exhibit contextuality but are bound states for all magic state distillation routines based on stabilizer codes of length less than $N$.\end{theorem}

This theorem, which is a consequence of a stronger theorem proven in section \ref{nogo}, places a universal limitation on qudit magic state distillation routines, but does not answer the second question. This is because it allows for an infinite sequence of magic state distillation routines, based on stabilizer codes of increasing length $N$, for which contextuality becomes sufficient in the limit $N \rightarrow \infty$.

\begin{figure}
\centering
\includegraphics[width=3.4in]{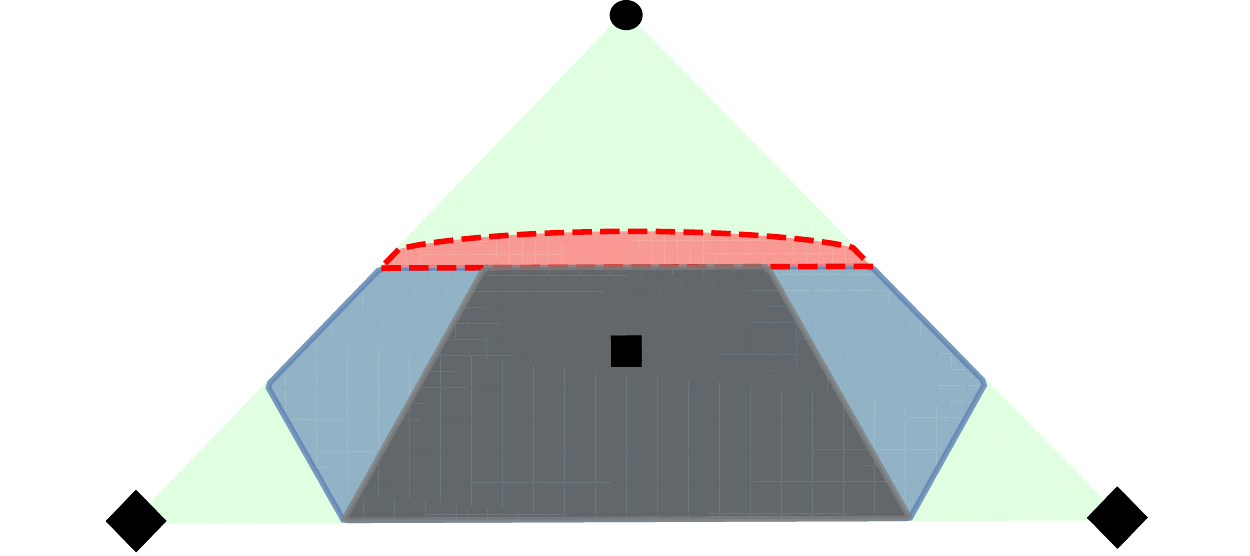}
\caption{\label{region}\textbf{Illustrating our theorem using a slice of qutrit state space.} The light green region denotes convex combinations of the three eigenstates of the Hadamard operator $H$ (represented by the black circle and black diamonds.) The blue region inside the light green region is the Wigner polytope, which contains the stabilizer polytope (dark grey shaded region) and the maximally-mixed state (black square.) For any magic-state distillation routine there exists a region of bound states near each face of the Wigner polytope, depicted as the red region with dashed border.}

\end{figure}

Our theorem is similar to, and inspired by \cite{bound} which demonstrated the existence of bound states for qubit magic state distillation outside the stabilizer octahedron. Our work differs from  \cite{bound} in two ways. Not only does our result apply to qudits; but, more importantly, we demonstrate the existence of bound states outside the Wigner polytope, which is the largest known classically simulable region of qudit phase space,\cite{Gross, Veitch_2012, PhysRevLett.109.230503} and contains the qudit stabilizer polytope as a proper subset. This includes bound states at a finite distance away from the stabilizer polytope, as illustrated in Figure \ref{region}. 

The remainder of this article is as follows. We first present a canonical form for qudit stabilizer reductions. We then reformulate magic state distillation in the language of discrete phase space and present  algorithms for simulating magic state distillation routines. Finally, we use these results to prove the existence of bound states for any finite stabilizer reduction.

\section{Qudit Stabilizer codes}
\label{notation}
Consider qudits of odd-prime dimension $d$. Generalized Pauli operators $X$ and $Z$ acting on qudits are defined as \cite{Gottesman1999}:
\begin{equation}
X \ket{k}=\ket{k+1}, ~ Z \ket{k}=\omega^k \ket{k}
\end{equation}
where $k \in \mathbb Z_d$ and $\omega=e^{2\pi i /d}$.

A multi-qudit Pauli operator acting on $N$ qudits,
\begin{equation}
P(\vec{a},\vec{b})= Z^{a_1} X^{b_1}\otimes Z^{a_2}X^{b_2} \otimes \ldots  \otimes Z^{a_N}X^{b_N},
\end{equation}  can be specified by a symplectic vector: $\left (\vec{a} | \vec{b} \right) \in \mathbb Z_d^N \otimes \mathbb Z_d^N $. Two multi-qudit Pauli operators $(\vec{a}|\vec{b})$ and $(\vec{a}'|\vec{b}')$ commute if
\begin{equation}
 \vec{a}\cdot \vec{b}' - \vec{b}\cdot \vec{a}' = 0.
\end{equation}

All known magic-state distillation routines are \textit{stabilizer reductions}, which consist of projecting the  $N$ input qudits onto the subspace of an $N$-to-$1$ qudit stabilizer code, possibly preceded by the random application Clifford unitaries to impose convenient symmetries on the input states. As argued in \cite{structure}, for any magic state distillation routine that is not a stabilizer reduction, there exists a stabilizer reduction with equal or better threshold fidelity.

A stabilizer code that encodes a single qudit in $N$ physical qudits is defined by a set of $N-1$ linearly-independent commuting $N$-qudit Pauli operators, whose symplectic vectors can be combined into a matrix: 
\begin{equation}
\mathbf{M}=
\begin{pmatrix}
\vec{a}_1^T  & | & \vec{b}_1^T \\
\vdots & | & \vdots \\
\vec{a}_{N-1}^T  & | & \vec{b}_{N-1}^T
\end{pmatrix}
= 
\begin{pmatrix}
\mathbf{\alpha} & | & \mathbf \beta
\end{pmatrix}.
\end{equation}

Two stabilizer codes are equivalent if they differ only by any combination of the following procedures:
\begin{enumerate}
	\item Exchange of qudit $i$ with qudit $j$, i.e., interchanging the $i$th and $j$th column of both the matrix $\mathbf \alpha$ and the matrix $\mathbf \beta$.
	\item Raising a stabilizer to a non-zero power in $\mathbb Z_d$, i.e., multiplying any row of  $\mathbf M$ by a non-zero element of $\mathbb Z_d$.
	\item Multiplying one stabilizer with another stabilizer, i.e., adding one row of $\mathbf M$ to another row. 
\end{enumerate}

Using these elementary operations, we can put any $(N-1)\times 2N$ matrix 
$\mathbf{M} =\begin{pmatrix}
\mathbf \alpha & | & \mathbf \beta
\end{pmatrix}$ into the following canonical form, following \cite{1997PhD},
\begin{equation}
\mathbf M_\text{canonical} = 
\begin{pmatrix}
\mathbf 1_{n\times n} & \mathbf A & \vec{A} & | & \mathbf B & 0_{n\times m} & \vec{B} \\
0_{m\times n} & 0_{m\times m} & 0_{m\times 1} & | & \mathbf C & \mathbf 1_{m \times m} & \vec{C}
\end{pmatrix}, \label{canonical}
\end{equation}
where $n$ is the rank of $\alpha$, and  $m=N-1-n$. $\mathbf A$ is an $n\times m$ matrix, $\mathbf B$ is an $n \times n$ matrix, and $\mathbf C$ is an $m \times n$ matrix. $\vec{A}$ and $\vec{B}$ are $n$ dimensional column vectors and $\vec{C}$ is an $m$-dimensional column vector.  The subspace where each stabilizer in the above canonical form has a given eigenvalue $\lambda_i=\omega^{s_i}$, defines the codespace.

We must also define logical Pauli operators, $X_L$ and $Z_L$, that satisfy $Z_L X_L = \omega X_L Z_L$ and commute with all stabilizers of the code. The allowed logical Pauli operators are parameterized by $(u,v) \in \mathbb Z_d \otimes \mathbb Z_d$.
\begin{equation}
(\vec{a}_L^T | \vec{b}_L^T) = u \begin{pmatrix} 0 & - \vec{C}^T & 1 | \vec{B}^T & 0 & 0 \end{pmatrix}+v \begin{pmatrix} 0 & 0 & 0  | -\vec{A}^T & 0 & 1 \end{pmatrix}.
\end{equation}

If $\vec{A}$, $\vec{B}$ and $\vec{C}$ are all zero, the code is trivial. Projecting onto the codespace of such a code is equivalent to simply projecting the first $N-1$ qudits onto some stabilizer state, and leaving the $N$th qudit unchanged.

\section{The Wigner polytope}
\label{Wigner}
We now review the definition of the Wigner polytope and its relation to contextuality.

\subsection{Discrete Wigner functions}
Given any $N$-qudit state $\rho$, we can define a quasi-probability distribution known as the discrete Wigner function $W^{(N)}_{\rho} (\vec{z},\vec{x})$ as in \cite{Wootters1987, Gross, Veitch_2012}. This is done by first defining \textit{phase-point} operators $A^{(N)}_{\vec{u},\vec{v}}$ :
\begin{eqnarray*}
A^{(N)}_{(\vec{0},\vec{0})} & = & \sum_{\vec{z},\vec{x}} \omega^{-2^{-1} \vec{z}\cdot \vec{x}} P(\vec{z},\vec{x}) \\
A^{(N)}_{(\vec{u},\vec{v})} & = & P(\vec{u},\vec{v}) A^{(N)}_{(\vec{0},\vec{0})} P^\dagger(\vec{u},\vec{v}),
\end{eqnarray*}
in terms of which, the Wigner function is defined as,
\begin{equation}
    W^{(N)}_\rho(\vec{z},\vec{x}) = \frac{1}{d^2}\text{Tr } \rho A^{(N)}_{(\vec{z},\vec{x})}.
\end{equation}
If the state is separable, $\rho=\rho_1 \otimes \rho_2 \otimes \ldots \otimes \rho_N $ the  $N$-qudit Wigner function  can be written as a product single-qudit Wigner functions:
\begin{equation}
W^{(N)}_{\rho} (\vec{z},\vec{x}) = \prod_{i=1}^N W^{(1)}_{\rho_i} (z_i, x_i).
\end{equation}

The set of single-qudit states with $W_\rho(z,x)\geq 0$ defines a convex set called the Wigner polytope. 

For Clifford operations and stabilizer measurements acting on states described by non-negative Wigner function, classical simulation \cite{Veitch_2012, PhysRevLett.109.230503} is possible thanks to the existence of an ontological model, in which the ``ontological state'' of the system is described by a single point in discrete phase space $(\vec{z}, \vec{x}) \in \mathbb Z_d^n \otimes \mathbb Z_d^n$. Clifford unitaries act as symplectic rotations and translations on the ontological state \cite{Appleby, Appleby2008}. Measurement of a multi-qubit Pauli operator $P(\vec{a},\vec{b})$ on a system in the ontological state $(\vec{z},\vec{x})$ yields the deterministic result $\omega^{\vec{a}\cdot \vec{x}-\vec{b}\cdot \vec{z}}$. It also acts as a random translation in the direction $(\vec{a},\vec{b})$. 

A quantum state $\rho$ within the Wigner polytope corresponds to a probability distribution over these ontological states given by its discrete Wigner function $W^{(N)}_\rho(\vec{z},\vec{x})$. If this probability distribution can be efficiently sampled, which is the case if $\rho$ is separable, then the above construction is an efficient scheme for classical simulation.

\subsection{Wigner negativity and contextuality}
\label{cont}
We review the equivalence established by \cite{nature} between Wigner negativity and contextuality with respect to stabilizer measurements. First, observe that contextuality implies negativity -- for any state with non-negative Wigner function, the construction of \cite{Veitch_2012}  reviewed above provides a non-contextual ontological model that reproduces stabilizer measurements. To show that negativity also implies contextuality, \cite{nature} construct a non-contextuality inequality that is violated by states with negative Wigner function.

To do this, we need to introduce an ancilla qudit, which can be in any state. Define $\{\Pi\}^{(u,v)}_{\text{sep}}$ to be the set of two-qudit projectors onto states of the form $\ket{\phi_{(u,v)}}\otimes \ket{k}$, where $\ket{\phi_{(u,v)}}$ is any stabilizer state whose discrete Wigner function does not have support on $(u,v)$. There are $d(d^2-1)$ such projectors. Next, define $\{\Pi\}_{\text{ent}}$ to be the set of rank-1 projectors onto two-qudit entangled stabilizer states. By the Choi-Jamiolkowski isomorphism, the number of such projectors equals the number of single-qudit Clifford unitaries, which is $d^3(d^2-1)$.

We now construct an exclusivity graph out of the combined set of projectors $V=\{ \Pi \}^{(u,v)}_{\text{sep}} \cup \{\Pi\}_{\text{ent}}$. Each projector corresponds to a vertex in the graph. Two vertices are connected by an edge if they correspond to orthogonal projectors. (Orthogonal projectors represent mutually exclusive measurement outcomes.) Figure \ref{exgraph} depicts the resulting exclusivity graph for qutrits, which contains $240$ vertices and $7116$ edges.

Define $\mathbf \Sigma^{(u,v)}$ to be the sum of all projectors in $V$. Compatibility with a non-contextual hidden variables (NCHV) model places an upper bound on $\langle \mathbf \Sigma^{(u,v)}\rangle$. In any ontological state, each projector (i.e., vertex) is assigned a value of $0$ or $1$, subject to the constraint that two orthogonal projectors (i.e., two adjacent vertices) cannot both be assigned the value $1$. Hence the set of all projectors assigned the value $1$ forms an \textit{independent set} in the exclusivity graph. The \textit{independence number} of a graph is the size of its largest independent set. In any ontological state or probabilistic-mixture of ontological states, $\langle \mathbf \Sigma^{(u,v)}\rangle$ cannot exceed the independence number of the exclusivity graph. 
The authors of \cite{nature} showed that the independence number of the exclusivity graph constructed from $V$ is $d^3$. Hence, any quantum state for which $\langle \mathbf \Sigma^{(u,v)}\rangle > d^3$ is incompatible with a NCHV model, and exhibits contextuality.

Using, e.g., \cite{cormick2006classicality}, the sum of projectors $\mathbf \Sigma^{(u,v)}$ can be rewritten as 
\begin{equation}\mathbf \Sigma^{(u,v)} = \left(d^3 \mathbf 1_{d\times d} - A^{(1)}_{(u,v)}\right) \otimes \mathbf 1_{d \times d}.
\end{equation} Therefore, the condition that $\text{Tr }  \rho A^{(1)}_{(u,v)}  <0$ is equivalent to $\text{Tr } \left(\mathbf \Sigma^{(u,v)} \rho \otimes \sigma \right) > d^3$, which is precisely the condition for contextuality obtained via the exclusivity graph construction.

\section{Simulating magic state distillation in discrete phase space}
\label{sim}

We now illustrate how finite stabilizer reductions can be simulated in discrete phase space. 

\subsection{Monte Carlo algorithm}
We first describe a Monte Carlo algorithm for simulating magic-state distillation, which illustrates the classical simulability of states within the Wigner polytope, in the spirit of \cite{Veitch_2012}. 

As in \cite{MSD}, we assume that the undistilled input qudits are all described by the same single-qudit density matrix $\rho_{\text{in}}$. Our algorithm,  takes as input: \begin{itemize}
    \item A single-qudit discrete Wigner function $W^{(1)}_{\text{in}}(z,x)$  corresponding to $\rho_{\text{in}}$; which is non-negative, and
    \item a stabilizer code $\mathbf M$, with eigenvalues $s_i$. 
\end{itemize} It returns as output: 
\begin{itemize}
    \item the single-qudit discrete Wigner function $W^{(1)}_{\text{out}}(z,x)$ corresponding to the density matrix $\rho_{\text{out}}$ of the distilled qudit.
\end{itemize}
The algorithm consists of the following steps:
\begin{enumerate}
    \item Initialize a histogram $h(z,x)=0$ for all $(z,x)$ in $\mathbb Z_d \otimes \mathbb Z_d$. 
    \item Repeat the following procedure many times: 
    \begin{enumerate}
\item Randomly choose a point $(\vec{z},\vec{x})$ in $\mathbb Z_d^N \otimes \mathbb Z_d^N$ from the probability distribution 
\begin{equation}
\bar{W}_{\text{in}}^{(N)}(\vec{z},\vec{x}) = \prod_{i=1}^N W^{(1)}_{\text{in}}(z_i,x_i).
\end{equation} 
Here, $\vec{z}=(z_1,z_2,\ldots, z_N)$ and $\vec{x}=(x_1,x_2,\ldots x_N)$. Because the probability distribution is separable, this can be done efficiently.
\item Check if this point is in the codespace, by checking if 
\begin{equation}
\mathbf M \begin{pmatrix} \vec{x} \\ -\vec{z} \end{pmatrix} = \begin{pmatrix} s_1 \\ \vdots \\ s_{N-1} \end{pmatrix} \label{code-space}
\end{equation}

\begin{enumerate}
\item If the point is not in the codespace, do nothing. 

\item If it is in the codespace, calculate $z_L$ and $x_L$:
\begin{eqnarray}
x_L & = &  \vec{b}_z\cdot \vec{z} - \vec{a}_z \cdot \vec{x} \\
z_L & = &  -\vec{b}_x\cdot \vec{z} + \vec{a}_x \cdot \vec{x}.
\end{eqnarray}
and increment $h(z_L, x_L) \rightarrow h(z_L,x_L)+1$.
\end{enumerate}
\end{enumerate}

\item After the step 2 has been repeated a sufficient number of times,  obtain the Wigner function of the distilled state by normalizing $h(z,x)$:
\begin{equation}
W^{(1)}_{\text{out}}(z,x) =  \frac{h(z,x)}{\displaystyle \sum_{u=0}^{d-1} \sum_{v=0}^{d-1} h(u,v)}. \label{normalize}
\end{equation}

\end{enumerate}

\subsection{Exact algorithm}

We now present an algorithm that calculates $W^{(1)}_{\text{out}}(z,x)$ exactly given $W^{(1)}_{\text{in}}(z,x)$. 

A naiive way to do this would be to loop over all $d^{2N}$ points in discrete phase space. For each point $(\vec{z},\vec{x})$, check it is in codespace: if it is, calculate $x_L$ and $z_L$, as before, and then update the histogram:
\begin{equation}
h(z_L,x_L)\rightarrow h(z_L,x_L) + \prod_{i=1}^{N} W^{(1)}_{\text{in}}(z_i,x_i). \label{updateW}
\end{equation} 
After completing this loop  normalize $h(z,x)$, via equation \eqref{normalize}, to obtain $W^{(1)}_{\text{out}}$.

The above algorithm loops over all $d^{2N}$ points in $N$-qudit phase-space. However, we only need to loop over the $d^{N+1}$ points in the codespace, which can be done explicitly, as we now show. 

The codespace consists of the set of phase-space points $(\vec{x},\vec{z})$ satisfying equation \eqref{code-space}. This equation represents $N-1$ equations in $2N$ unknowns -- hence the space of solutions is an $N+1$-dimensional vector space. Assume each $s_i=0$. Then, the $N-1$ stabilizers of $M$ along with the logical $Z$ and logical $X$ operators are $N+1$ independent solutions to equation \eqref{code-space}, and form a basis for the codespace. We thus have:
\begin{equation}
\begin{pmatrix} \vec{z}^C(\vec{u},z_L,x_L) \\ \vec{x}^C(\vec{u},z_L,x_L) \end{pmatrix} = \begin{pmatrix}\mathbf{M}^T & \begin{matrix}\vec{a}_z  \\ \vec{b}_z \end{matrix} &  \begin{matrix}\vec{a}_x  \\ \vec{b}_x \end{matrix} 
\end{pmatrix} \begin{pmatrix}\vec{u} \\ -z_L \\ -x_L \end{pmatrix}. \label{codespace}
\end{equation}
$\vec{u}$ is an $N-1$ dimensional vector, which, along with $z_L$ and $x_L$, parameterizes the codespace. If $s_i \neq 0$ we would need to add a constant vector representing any particular solution to equation \eqref{code-space}, that can be obtained using a generalized inverse of $\mathbf{M}$, to the RHS of this equation. 

Explicitly, the algorithm is:
\begin{enumerate}
    \item Initialize a histogram $h(z,x)=0$ for all $(z,x)$ in $\mathbb Z_d \otimes \mathbb Z_d$. 
    \item For each $(z_L,x_L) \in \mathbb Z_d \otimes Z_d$, and each $\vec{u} \in \mathbb Z_d^{N-1}$: 
    \begin{enumerate} \item calculate $\vec{z}^C$ and $\vec{x}^C$ via equation \eqref{codespace}, and, 
    \item update $h(z,x)$ via equation \eqref{updateW}.
    \end{enumerate}
    \item Obtain $W^{(1)}_\text{out}(z,x)$ from $h(z,x)$ using equation \eqref{normalize}.
\end{enumerate}

This algorithm works even when $W^{(1)}_{\text{in}}$ contains negative entries. In fact, this formulation of magic state distillation is more general than the usual formulation, since it also applies to ``post-quantum" theories that allow preparations with discrete Wigner functions that do not correspond to quantum states.

\section{Impossibility Theorem}
\label{nogo}
With the above ingredients in place, we prove our theorem. We first prove a special case of our main result, which we then generalize. 

A natural candidate for a magic state is given by the Wigner function:
\begin{equation}
W(z,x) = \begin{cases}  \nu & (z,x)=(0,0) \\ (1-\nu)/(d^2-1) & (z,x) \neq (0,0). \end{cases}
\end{equation}
where $\nu < 0$. An example of such a state is $\ket{H_i}=\frac{1}{\sqrt{2}} \left( \ket{1}-\ket{2} \right)$ for qutrits \cite{ACB} for which $\nu=-1/3$. Because the discrete Wigner function has only one negative entry, this state lies directly above a single face tof the Wigner polytope, and is therefore analogous to the $\ket{T}$ state of \cite{MSD}.

Consider the state $\rho$ obtained from mixing this state with the maximally-mixed-state. By randomly applying a symplectic rotation \cite{Appleby}, any state can be put in this form. For such states, $W_\rho(0,0)=\nu$ is the only entry of the Wigner function which may be negative.  Any magic state distillation procedure induces a continuous function $\nu_{\text{out}}=f(\nu_{\text{in}})$.

\begin{figure}
\centering
\includegraphics[width=3.4in]{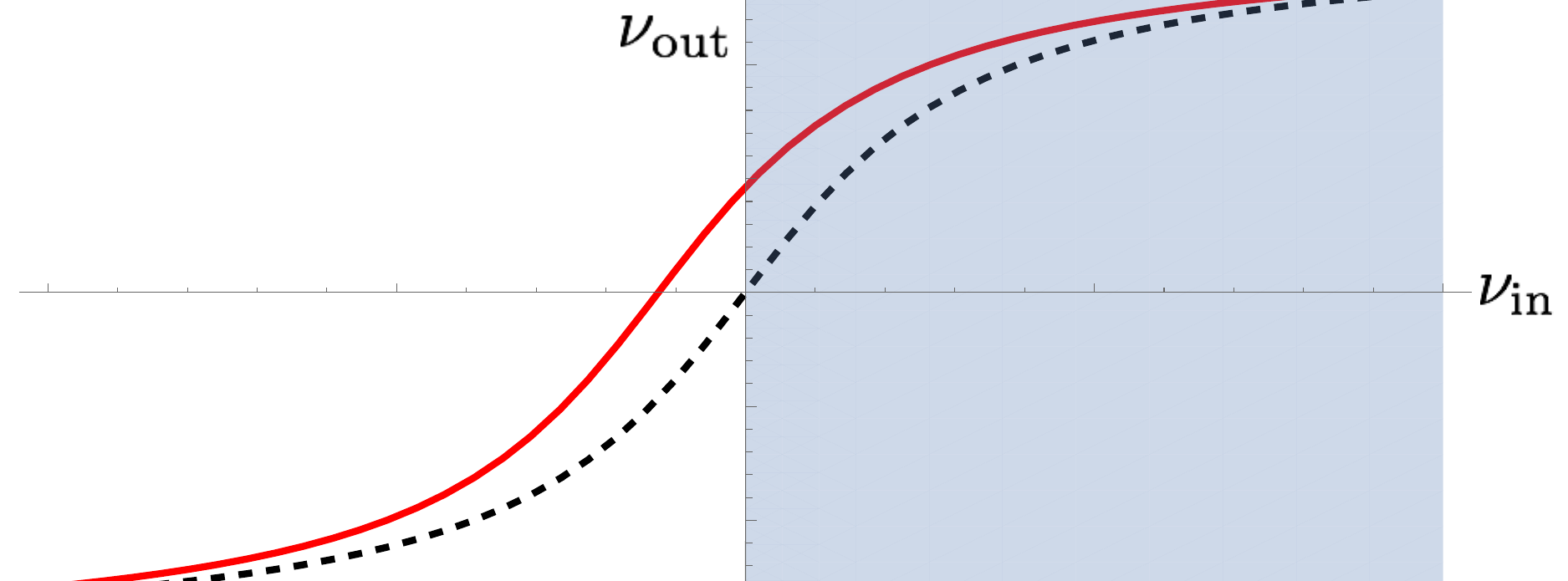}
\caption{\label{tight} \textbf{Schematic plots of the function $\nu_\text{out}=f(\nu_\text{in})$ induced by putative magic-state distillation routines.} The shaded region is the interior of the Wigner polytope. A tight magic-state distillation routine would induce a function $f$ which satisfies $f(0)=0$ (dashed black curve). We show that all finite magic-state distillation routines induce functions for which $f(0)>0$ (solid red curve).}
\end{figure}

By the usual argument of classical simulability, if $\nu_{\text{in}}>0$ then $\nu_{\text{out}}>0$ for any stabilizer reduction. For a stabilizer reduction with threshold tight to the Wigner polytope, $\nu_{\text{out}}<0$ if $\nu_{\text{in}}<0$. Therefore, $f(0)=0$ for any tight stabilizer reduction, illustrated in Figure \ref{tight}.

Equation \eqref{codespace} leads to an explicit expression for $\nu_{out}$ when $\nu_{in}=0$:
\begin{equation}
\begin{split}
\nu_{out}  & =W_{\text{out}}(0,0) = \sum_{\vec{u}} \bar{W}^{(N)}_{\text{in}}(\vec{z}^C(\vec{u}, 0, 0),\vec{x}^C(\vec{u}, 0, 0))/Z \\ & = \sum_{\vec{u}} \prod_i W^{(1)}_{\text{in}}\left(z^C_i(\vec{u},0,0), x^C_i(\vec{u},0,0)\right)/Z,
\end{split}
\end{equation}
where $Z$ is a (positive) normalization constant.
All entries of $W^{(1)}_{\text{in}}$ are positive except one: $W^{(1)}_{\text{in}}(0,0)=0$. The only way $\nu_{out}$ can be equal to zero is, if, for each $\vec{u} \in \mathbb Z_d^{N-1}$, there exists at least one value of $i$ such that both $z^C_i(\vec{u},0,0)=0$ and $x^C_i(\vec{u},0,0)=0$. 

Denote $\vec{u} = \begin{pmatrix} \vec{w} \\ \vec{v} \end{pmatrix}$ where $\vec{w}$ and $\vec{v}$ are $m$ and $n$ dimensional vectors. Using the canonical form of the stabilizer code, we have:
\begin{equation} \begin{split}
\begin{pmatrix} \vec{z}^C(\vec{u},0,0) \\ \vec{x}^C(\vec{u},0,0) \end{pmatrix}
 = 
\begin{pmatrix} \vec{w} \\ A^T \vec{w} \\ \vec{A}^T \vec{w} \\ \hline B^T \vec{w}+C^T \vec{v} \\ \vec{v} \\ \vec{B}^T \vec{w}+C^T \vec{v} \end{pmatrix}
\end{split}
\end{equation}

Consider a vector $\vec{u}$ for which $\vec{w}$ and $\vec{v}$ contain all non-zero entries. For such a vector, the only value of $i$ for which both $z^C_i$ and $x^C_i$ could equal $0$ is the last value. For this we require that both $\vec{A}^T \vec{w}=0$ and $\vec{B}^T \vec{w}+\vec{C}^T \vec{v}=0$. 

For prime $d\geq 3$, if $\vec{A}$ has a single non-zero entry, we can always find a vector $\vec{w}$ with all non-zero entries such that $\vec{A}^T \vec{w} \neq 0$. Similarly, if either $\vec{B}$ or $\vec{C}$ has a non-zero entry, we can find a pair of vectors $\vec{w}$ and $\vec{v}$, each with all nonzero entries, such that $\vec{B}^T \vec{w}+\vec{C}^T \vec{v}\neq 0$. 

Therefore, for $\nu_{out}=0$ we require $\vec{A}=0$, $\vec{B}=0$ and $\vec{C}=0$. But, as mentioned earlier, this means that the magic state distillation routine is trivial, and $\nu_{out}=\nu_{in}$. Thus, there is no finite stabilizer reduction that distills states tight to the $A_{(0,0)}$ face of the Wigner polytope.

We now generalize the above argument. First, restrict attention to states much closer to one facet than they are to any other facet. For this family of states we can unambiguously define $\nu$ to be the entry of the Wigner function smallest in magnitude.  Any magic-state distillation routine whose threshold is tight or almost tight to the Wigner polytope defines a continuous map on the space of states  $\rho_{\text{out}}(\rho_{\text{in}})$ which induces a continuous function $\nu_{\text{out}}=f(\nu_{\text{in}})$ of $\nu$. This function must have a fixed point at $\nu=0$, if the magic-state distillation routine has no bound states outside the Wigner polytope.

Our argument that $f(0)>0$ above made no use of the location of the single zero of the discrete Wigner function $W^{(1)}_{\text{in}}$, nor of its symmetries; so it generalizes to the following statement. Let $M$ be any finite stabilizer reduction. $M$ maps any state described by a non-negative discrete Wigner function with a single zero to a state described by a discrete Wigner function with all positive entries. In other words, $M$ maps states which are on a single face of the Wigner polytope to states in the interior. By continuity, there exist states just outside the face of the Wigner polytope which are also mapped to the interior. We thus have the following theorem:
\begin{theorem} \label{main2}
 Let $\rho$ be any state whose discrete Wigner function contains exactly one zero. For any finite magic state distillation procedure, all states sufficiently close to $\rho$, including states outside the Wigner polytope, are bound states. 
 \end{theorem}

\section{Outlook}
\label{outlook}

We have yet to understand the implication of our result for contextuality and Wigner negativity as a resource \cite{ Veitch_2012, PhysRevLett.109.230503, Veitch_2014, PhysRevLett.115.070501, Wang_2019}. In particular, what is the computational power of a Clifford computer with access to a large, but finite, supply of states which exhibit a small violation of a single non-contextuality inequality?\footnote{We thank Aram Harrow and Dan Browne for raising this question.} 

The result of \cite{nature} has been generalized to qudits of any odd dimension \cite{Delfosse_2017}, and it should be possible to extend our theorem to this more general case. 

Our theorem rules out the existence of finite magic-state distillation routines for which contextuality is sufficient for magic-state distillation. However, it allows the possibility that the threshold for magic state distillation can be made arbitrarily close to the the Wigner polytope by increasing the size of the stabilizer code used.  If this is the case, for any state $\rho$ exhibiting contextuality, there would exist an $N$-qudit magic-state distillation routine (with $N$ sufficiently large) for which $\rho$ is not a bound state; and in this sense, contextuality may still turn out be sufficient for universal quantum computation.

\begin{acknowledgements}
SP thanks Prof. P.S. Satsangi for inspiration and suggesting this problem to him. SP also thanks Mark Howard for comments on a draft of this manuscript.
The research of SP is supported in part by a DST INSPIRE Faculty award, DST-SERB Early Career Research Award (ECR/2017/001023) and MATRICS grant (MTR/2018/001077). Aashi Gupta acknowledges the support of an AADEIs Undergraduate Research Award.
\end{acknowledgements}

\bibliography{qudit}

\end{document}